\documentclass[final,5p,times,twocolumn,english,fleqn]{elsarticle}
\pdfoutput=1
\usepackage{amssymb}
\usepackage{babel}
\usepackage{graphicx}
\usepackage{amsmath,amssymb,amsfonts,threeparttable}
\usepackage{textcomp}
\usepackage{gensymb}
\usepackage{booktabs}
\usepackage{hyperref}
\hypersetup{
    breaklinks=true,
    unicode=false,          
    pdftoolbar=true,        
    pdfmenubar=true,        
    pdffitwindow=false,     
    pdfstartview={FitH},    
    pdfauthor={Venkat},     
    pdfsubject={ABL},   
    pdfcreator={Venkat},   
    pdfproducer={Venkat}, 
    pdfnewwindow=true,      
    colorlinks=true,       	
    linkcolor=black,          
    citecolor=black,    	    
    filecolor=black,   		    
    urlcolor=blue            
}
\journal{Journal of Magnetism and Magnetic Materials}
\usepackage[labelfont=bf]{caption}
\makeatletter
\def\ps@pprintTitle{%
 \let\@oddhead\@empty
 \let\@evenhead\@empty
 \def\@oddfoot{}%
 \let\@evenfoot\@oddfoot}
\makeatother

\makeatletter
\def\blfootnote{\gdef\@thefnmark{}\@footnotetext}
\makeatother

\begin{document}
 \title{Magnetization spin dynamics in a (LuBi)$_3$Fe$_5$O$_{12}$ (BLIG) epitaxial film}
 \author[label1]{M. Malathi\corref{mycorrespondingauthor}}
 \cortext[mycorrespondingauthor]{Corresponding author}
 \ead{malathi.fil@ee.iitm.ac.in}
 \author[label1]{G. Venkat}\author[label1]{A. Arora} \author[label2]{I. I. Syvorotka} \author[label3]{V. Sivasubramanian}
 \author[label1]{A. Prabhakar} 
 \address[label1]{Dept. of Electrical Engineering, Indian Institute of Technology Madras, India 600036.} \address[label2]{Department of Crystal Physics and Technology, Scientific Research Company "Carat", Lviv, Ukraine.}\address[label3]{Condensed Matter Physics Division, Indira Gandhi Centre for Atomic Research, HBNI, Kalpakkam - 603 102, India.}

 \begin{abstract}

 Bismuth substituted lutetium iron garnet (BLIG) films exhibit larger Faraday rotation, and have a higher Curie temperature than yttrium iron garnet.
 We have observed magnetic stripe domains and measured domain widths of
 1.4 $\mu$m using Fourier domain polarization microscopy, Faraday rotation 
experiments yield  a coercive field of 5 Oe. These characterizations form the 
basis of micromagnetic simulations that allow us to estimate and compare 
spin wave excitations in BLIG films. We observed that these films
 support thermal magnons with a precessional frequency of 7 GHz with a
 line width of 400 MHz. Further, we studied the dependence of 
precessional frequency on the externally applied magnetic field. Brillouin light
 scattering experiments and precession frequencies predicted by simulations show similar trend  
 with increasing field.

\end{abstract}


\maketitle

\blfootnote{BLS - Brillouin light scattering spectroscopy; SW - Spin Wave; FR - Faraday rotation;  BLIG - Bismuth Lutetium Iron Garnet;
PM - Polarization microscopy; MOFE - Magneto optic Faraday effect}
\section{Introduction}\label{intro}
Spin waves also known as magnons,  have been extensively explored in the past decade for
 a variety of magnetic devices like multiplexers, logic gates, waveguides and resonators
 \cite{Urazhdin,KVogt,Kostylev,Krawczyk,Stamps}. There have been recent experimental
 demonstrations of magnetic domain walls as re-configurable nano sized magnonic waveguides \cite{Wagner16}. 
The preferred choices of ferromagnetic materials for these devices are permalloy and
 CoFeB. Another class of popular materials include insulating ferrimagnetic materials, or
 ferrites, like yttrium iron garnet (YIG) and bismuth substituted
lutetium iron garnet (BLIG). 
We know that the spin wave decay length in permalloy is 3 orders shorter than that of
ferrites. BLIG ((LuBi)$_3$Fe$_5$O$_{12}$) also exhibits a higher Faraday rotation,  
at a higher Curie temperature, than YIG. This makes BLIG films better
candidates for use in magneto-optic devices \cite{Syvorotka12}.
\par Ferrite films can be used for novel applications such
as magneto-optic Q switching \cite{Goto:16}. They have been used in a wide variety
of other applications \cite{YIG_Magnonics} and are also being considered for demonstrating logic
devices such as majority gates with fast clocking frequencies \cite{ganzhorn2016magnon}.
\par In this work, we characterize an optically transparent BLIG film of thickness 7.9 $\mu$m,  
epitaxially grown over a gadolinium gallium garnet
(GGG) substrate \cite{Syvorotka12}, \cite{Syvorotka_OSA}. We first measure
the magneto-static and magneto-optic properties of these films, using external magnetic fields. 
We then use the parameters extracted from the static analysis in the micromagnetic
simulations to study the dynamic properties of the thermally excited spin waves.
Finally, we use Brillouin light scattering (BLS) experiments to corroborate the
predictions of micromagnetic simulations. For measurement of Faraday rotation and coercivity, we designed and used 
the Magneto optic Faraday effect (MOFE) experiment discussed in section \ref{FR}. In section \ref {PM}, we characterize 
striped domain patterns observed using a transmission mode polarization
microscope (PM). Degaussing the film, 
by applying alternate positive and negative decreasing magnetic
fields, produces orderly striped domain patterns. When observed in the Fourier plane of the microscope, these orderly
domains show a diffraction pattern similar to a 1D grating. \par  In section \ref{simulation}, we extract the properties of BLIG 
from the previous static measurements and simulate the multiple stripe domains using the micromagnetic solver MuMax3 \cite{MuMax3}.
We allow the simulations to relax to a ground state where the film has a domain width, which is decided by the initial magnetization. 
To the best of our knowledge, dynamic micromagnetic simulations of magnetic
oscillations in stripe domain structures have not been reported elsewhere. 
\par We begin with a theoretical estimate of parameters like magnon frequency and domain wall frequency. 
We then include thermal fluctuations and external magnetic fields to the static
 multi domain simulation model for the dynamic analysis.
Brillouin light scattering (BLS) measurements, at room temperature, confirm 
our estimated increase in magnon frequency with an increase in external 
magnetic field, as discussed in section \ref{Dynamic}.

\section{Hysteresis measurements}\label{FR}
\begin{figure}[th]
   \centering{}
    {
     \includegraphics[width=1\columnwidth]{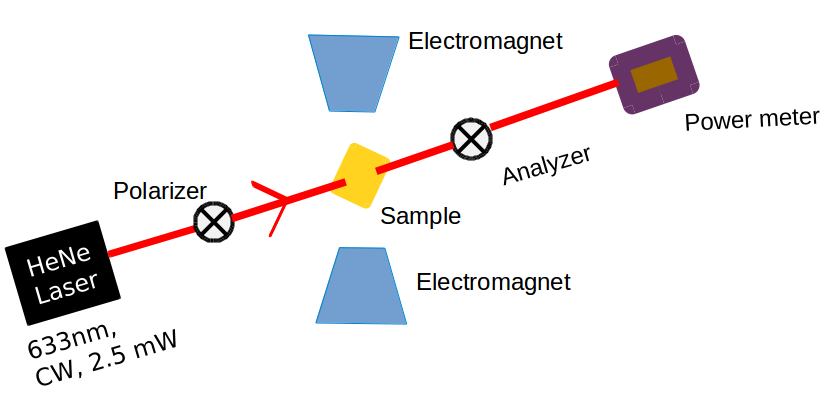}
     \protect\caption{ \label{fig:Fig-1} MOFE experimental setup to measure Faraday rotation}
    }
  \end{figure}

The MOFE set-up shown in Fig.~\ref{fig:Fig-1}, consists of a He-Ne laser ($\lambda=633$ nm) with an average power of 2.5 mW,
a polarizer and an analyzer, and a power meter with a measurement range of 50 nW - 50 mW.
We use a pair of electromagnets with a maximum magnetic field of $\pm 600\,\text{Oe}$ to provide the external magnetic field.  
The procedure to obtain the Faraday rotation $\left(\theta_{\text{F}} \right)$ from the film as a function of the applied magnetic field
$\left(\mathbf{H}_{\text{app}} \right)$ is as follows:
\begin{itemize}
  \item We first obtain the change in transmitted optical power $\left(P_{\text{tr}} \right)$ as a function of
	analyzer angle $\left(\theta_{\text{an}} \right)$. This will follow Malus' law
	\begin{equation}
	 P_{\text{tr}} = P_{0} \cos^{2}\left(\theta_{\text{an}}+\phi \right),\label{eq:1} 
	\end{equation}
	The values of $P_{0}$ and $\phi$ account for any misalignment and absorption that occur in the setup.
	The variation in transmitted power is shown in Fig.~\ref{fig:Fig2a}.
	\begin{figure}[th]
   \centering{}
    {
     \includegraphics[width=0.75\columnwidth]{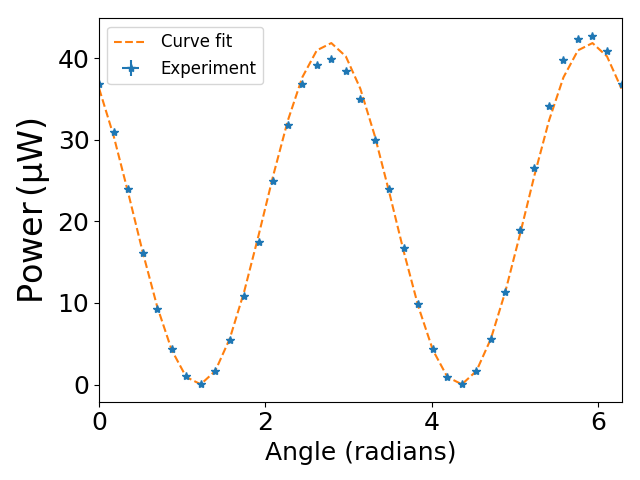}
     \protect\caption{ \label{fig:Fig2a} Transmitted optical power as a
                     function of rotation of analyzer angle.
                      The dashed line shows a fit to (\ref{eq:1}).
                      }
     }
  \end{figure}
    \item We then measure the variation of $P_{\text{tr}}$ with an applied magnetic field and map this to a
 	variation in Faraday rotation angle using (\ref{eq:1}). This variation is shown as the hysteresis loop in 
	Fig.~\ref{fig:Fig-3}.
   \end{itemize}
The film shows low coercivity of
$\sim 5\,\text{Oe}$, which indicates that the magnetization lies in the
plane of the film. The film also shows a low remanence of
15 \% of $M_{\text s}$  which is consistent with soft magnetic materials.
\begin{figure}[th]
   \centering{} {
     \includegraphics[width=0.75\columnwidth]{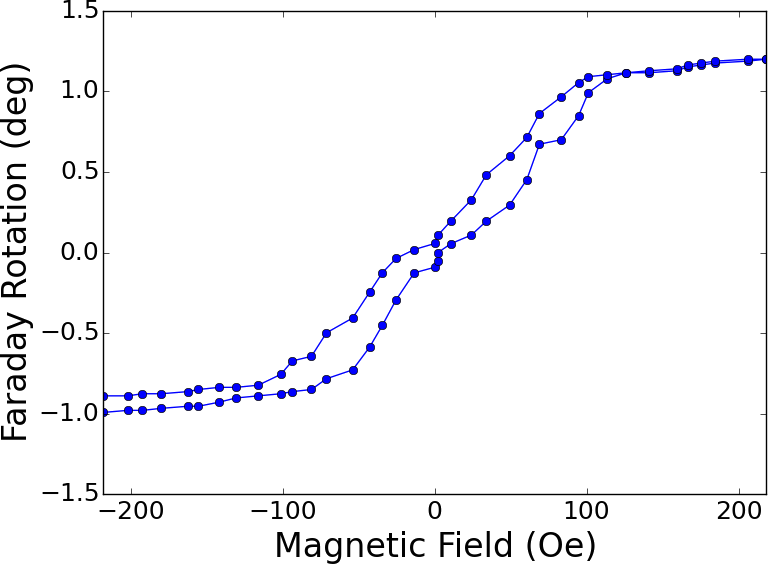}
     \protect\caption{ \label{fig:Fig-3} Hysteresis loop of the film
       obtained by Faraday rotation.} }
  \end{figure}
\section{Domain imaging using polarization microscopy (PM)}\label{PM}
\par
\begin{figure}[th]
\centering{} { \includegraphics[width=1.0\columnwidth,height=0.37\columnwidth]{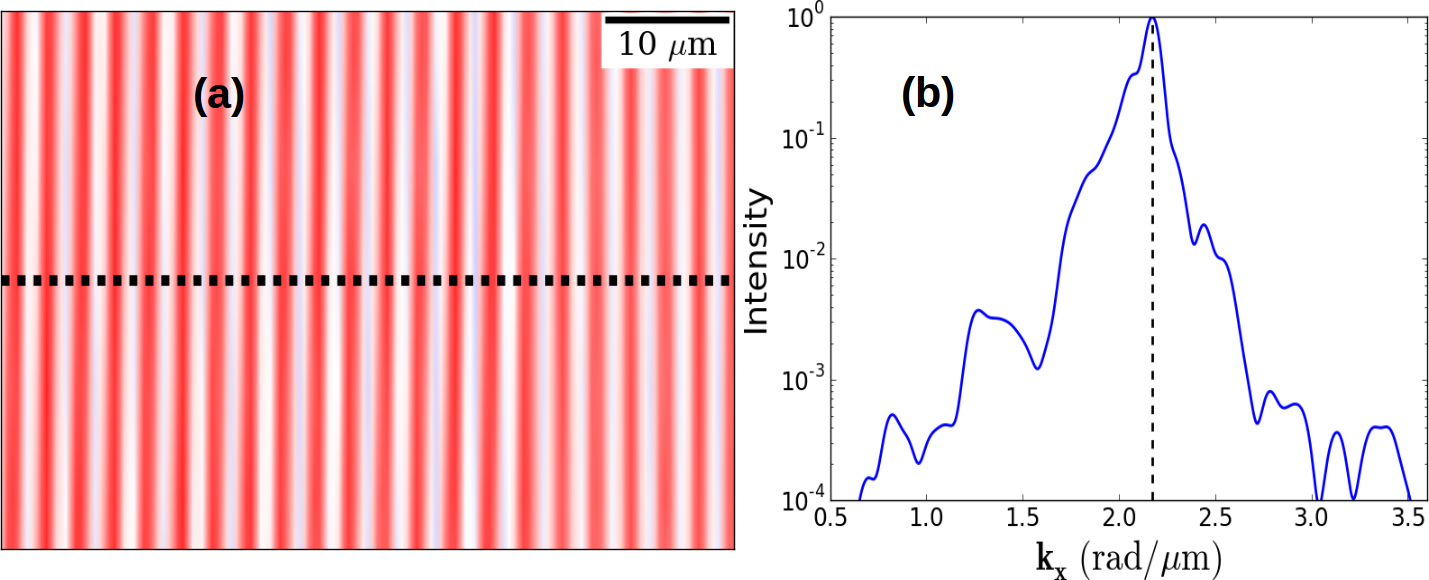}
  \protect\caption{ \label{fig:Fig-4} (a) Domain images of the
    sample obtained by polarization microscopy. Red and white
    regions correspond to alternating domain patterns. (b)
    Fourier transform along the black dotted line in (a). The peak corresponds to a domain period of
    $2.8\,\mu\text{m}$ and a domain width of $1.4\,\mu\text{m}$.}}
\end{figure}

 We use polarization microscopy to observe the spatial orientation of the magnetic domains in these ferrite films.
The setup consists of a transmission mode optical microscope with two linear polarizers 
in a cross axis polarizer-analyzer configuration. The light from a tungsten halogen lamp, after passing through the 
polarizer is incident on the sample and collected using a 20x objective. The analyzer, which is in a cross axis
 configuration placed after the objective in the optical path, 
 diminishes the direct transmission and obtains the dark field images which is captured by a CCD camera.
 In Fig.~\ref{fig:Fig-4} the images obtained show stripes of alternating intensity patterns attributed to 
domains of opposite in-plane magnetization, resulting from a differential Faraday effect.
\par We observed uniformly magnetized domains along the in-plane easy axes. With the
film thickness in microns, domains should be separated by straight parallel Bloch walls.
In the absence of any externally applied field, and when
the film is demagnetized, we observed that the total volumes of the two sets of
domains were equal, and the walls were equally spaced. 
The spacing of the walls is governed by a host of competing factors. 
The exchange and anisotropy energies of the walls favour wide domains 
while the dipolar energy of the domains favours closely spaced walls. 
The spacing of the walls is a compromise between these effects to achieve minimum energy state \cite{Chung75}. 
This was evident on the application of a magnetic field to the demagnetized films in a direction parallel to the domains. 
With an increase in magnetic field, each domain experiences a
torque that tends to turn it in the direction of the applied field. 
As a result, the exchange energy increases and domains grow wider, 
finally becoming single domain at $\approx$100 Oe. 

\par We can infer the average domain width from PM images using a Fourier spectrum of the
scan along the black dotted line in Fig.~\ref{fig:Fig-4}(a). The Fourier spectrum of the scan in Fig.~\ref{fig:Fig-4}(b), 
shows a single peak giving us a period of domain $\lambda_{\text{DW}}=2.8\,\mu\text{m}$. Since one period in
Fig.~\ref{fig:Fig-4}(a) consists of oppositely aligned domains, this
gives us a domain width of $\delta~ =\, 1.4\,\mu\text{m}$. 

\begin{figure}[th]
\centering{} {\includegraphics[width=0.9\columnwidth]{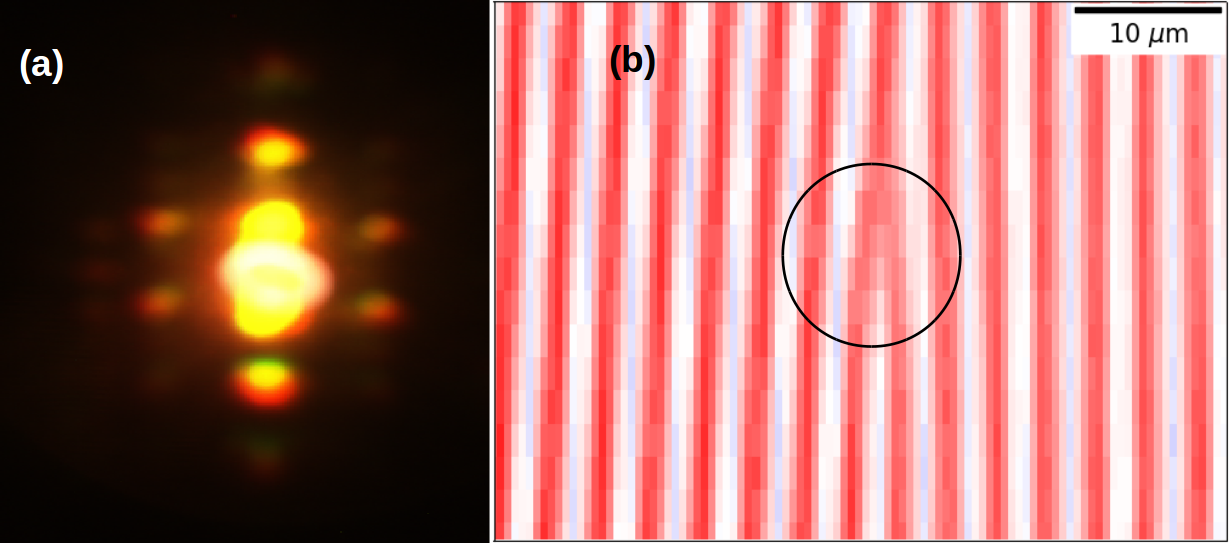}
  \protect\caption{ \label{fig:Fig-5}(a) Fourier plane image of
    the domain images obtained by inserting a lens in the ray path in
    the microscope.  The different spots correspond to diffraction
    orders due to a magneto-optical grating formed by the domains.
    (b) The 'fork' domain patterns obtained in the film indicate a change in the alignment of the magnetization.}}
\end{figure}

\par The Fourier domain image of the sample, as shown in
Fig.~\ref{fig:Fig-5}(a) is viewed by
placing a concave lens in the optical path of the microscope after the analyzer. 
The multiple spots correspond to diffraction
orders from the magneto-optical grating formed by the alternating
domains which are similar to the diffraction patterns obtained from bi-periodic
stripe domains \cite{Arzamastseva2006}. We observed a rotation of the diffraction orders on changing 
the direction of magnetic field which has been discussed elsewhere \cite{Diff-spots-rotation}. 
We also observe 'fork' domain patterns shown in Fig.~\ref{fig:Fig-5}(b) in our
films, which mark regions where the alignment of the domains changes
significantly. These alignment changes may happen due to a variety of
reasons such as structural deformities or local stray field
variations. The analysis of these fork domains and their application is an ongoing work.


\section{Multi domain simulations}\label{simulation}
\begin{figure}[th]
    \centering{} {
      \includegraphics[width=0.6\columnwidth]{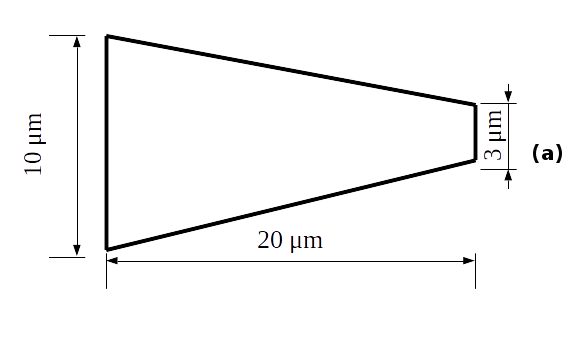}
      \includegraphics[width=1.0\columnwidth]{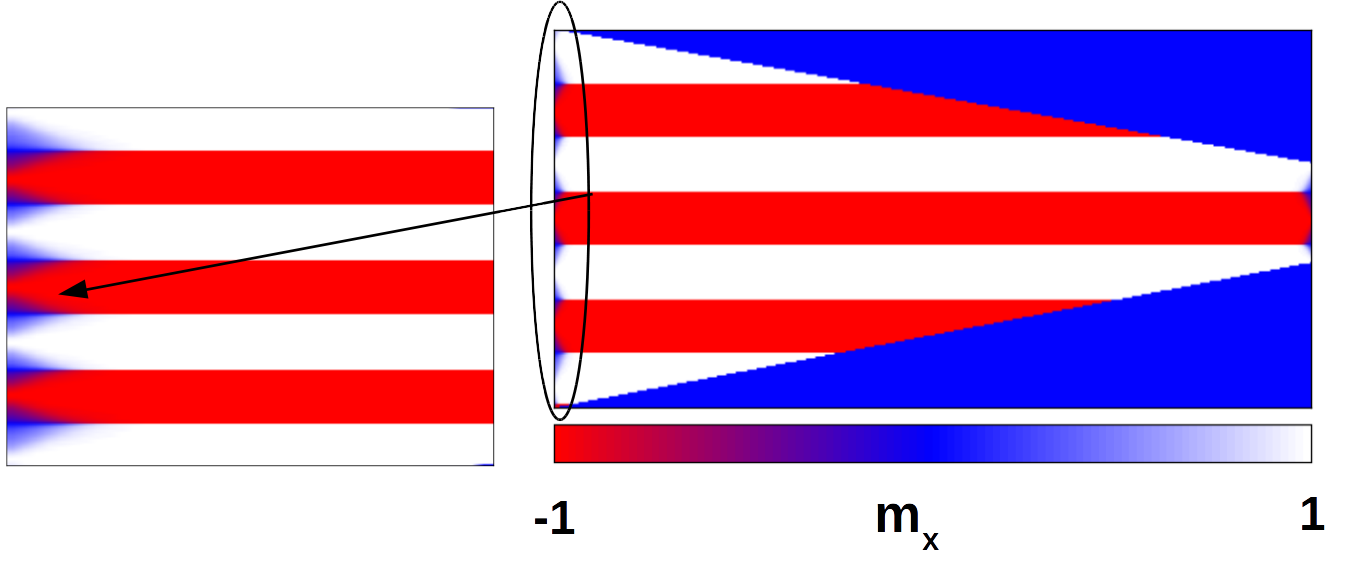}
      \protect\caption{ \label{fig:Fig6}(a) A schematic of the
        structure under study. The thickness was kept at
        $7.9\,\mu\text{m}$. (b) The ground state of the film
        showing multiple domains along the width of the sample. The
        edges show flux closure domain states which are visible in the
        zoomed inset figure.}}
  \end{figure}
The film, shown in
Fig.~\ref{fig:Fig6}(a), has the same shape and thickness as that of the BLIG film with the in-plane dimensions scaled to ensure that the simulation
does not become computationally prohibitive. We use a GPU accelerated micromagnetic
package MuMax3 to simulate the multi domain state. The cell size chosen for the
simulation is $50~\times~50~\times~50\,\text{nm}^3$, that is
higher than the exchange length of BLIG $\left(13\,\text{nm}
\right)$ but is sufficient to resolve the magnetization dynamics of the structure. A further reduction in cell edges, to 45 nm, did not affect the magnon frequency.
We consider the parameters of BLIG to be: saturation magnetization
($M_{\text{s}}$) as $1.27 \times 10^{5}$ A/m \cite{Syvorotka12}, uniaxial anisotropy
constant ($K$) as $5.9\times10^{3}$~J/m$^3$ \cite{denysenkov_99}, exchange constant
($A)$ as $4\times10^{-12}$ J/m, \cite{dadoenkova_16}. 
\par At the beginning of the simulation, the magnetization in the system 
is set in a stripe domain state and then allowed
to relax. In the relaxed ground state, shown in
Fig.~\ref{fig:Fig6}(b), we observe that the striped domain state is still preserved and that 
flux closure magnetization states appear close to the edges. This is due to higher 
demagnetization fields close to these edges. We measure a domain width  $\delta_{\text{sim}} = 1.4\,\mu\text{m}$,
approximately the same as that reported in section \ref{PM}. We also repeated the simulations with a different initial
conditions to obtain domain widths of  $ 0.45\,\mu\text{m}$. 

\section{Dynamic analysis of BLIG films}\label{Dynamic}
 \par Dipolar magnons are observed to have
wavelengths between several microns and millimeters \cite{Sandweg10}
while the localized thermal exchange magnons have wavelengths in
nanometers. The frequency of these exchange magnons depends very
strongly on the local magnetization. Thermally induced magnetization
gradients were studied using BLS measurements of the local frequency
\cite{Agrawal13}. The orientation of the magnetization in each
domain, and the domain size, are determined by the balance between
crystal anisotropy, magnetoelastic energy, domain wall energy and
magnetostatic energy.

\subsection{Frequency of oscillation of magnons}
\par A magnetic sample when incident with a laser beam of power 10 - 20 mW has negligible local heating
and this cannot affect the magnetization dynamics. However, the incident photons are inelastically
scattered due to thermal magnons which exist even at room temperature.
Light scattering from thermal magnons in YIG have been observed earlier using visible laser light \cite{Sandercock73}. 

The wavevector of these magnons, in a backscattering geometry, is given by, $k_{\rm m}\ =\ 4\pi n/\lambda_{0}$, 
where $n$ is the refractive index of
the material at the wavelength $\lambda_{0}$ of the incident light. For calculations, we used $n=2.44$ at $\lambda_{0} = 532$ nm,
(used in experiments later) \cite{Doormann}. We calculated the wavevector of the
magnons, $k_{\rm m}\ =\ 57\ \mu\rm{m}^{-1}$ and the corresponding wavelength $\lambda_{\rm m}=110$ nm, which indicates that these are thermal
exchange magnons. The dispersion relation of these magnons in the exchange limit is given by \cite{Anil_99},
\begin{equation}
\omega_{\rm{MAG}}^2 = \omega_{0}(\omega_{0}+ \omega_{\rm{M}}) 
\end{equation}
where,
\begin{eqnarray}
\omega_{0} &=& \gamma\mu_{0}(H_{0}+H_{\rm{ani}}+H_{\rm{demag}}+H_{\rm{exch}})\ \rm{and}\nonumber\\
\omega_{\rm{M}}&=&\gamma\mu_{0}M_{\rm s}\nonumber
\end{eqnarray}

$H_{0},~H_{\rm{ani}},~H_{\rm{demag}}$ and $H_{\rm{exch}}$ are the applied magnetic field, 
anisotropy field, demagnetization field and exchange field respectively. In the absence of an external magnetic field,
i.e., $H_{0}~=~0$, we obtain the magnon frequency of 8.4 GHz.

  \subsection{Frequency of oscillation of domain walls}\label{domains}

\par At room temperature, domain walls can also oscillate due to the thermal energy ($E_{\rm {Th}} =  77.6$ meV).
The nature of these oscillations is determined by an effective mass density of the wall and a restoring force
generated by an externally applied field and demagnetization fields. The domain wall mass density is a property
of the structure of the moving domain wall. The domain wall resonance frequency is given by \cite{Djiles}
\begin{equation}
\omega_\text{DWR}= \sqrt{\frac{\Gamma\mu_o M_{\rm{s}}^2A}{m\chi_{\rm{in}}}}
\end{equation}
where, $\Gamma = \frac{4}{3}$ for parallel Bloch walls, $m$ is Bohr magneton, and
the initial susceptibility $\chi_{\rm{in}} = 4.25$. Using these values, we obtain $\omega_{\text{DWR}} = 46$ MHz,
which is approximately an order less than the thermal magnon frequency.
\subsection{Dynamic simulation}
\begin{figure}[tbh]
   \centering{} {\includegraphics[width=0.9\columnwidth]{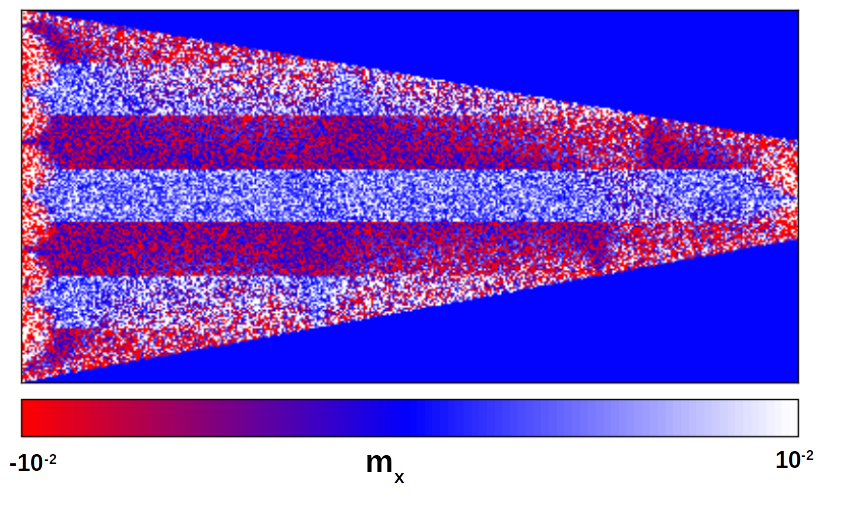}
     \protect\caption{ \label{fig:Fig-7} Room temperature driven
       fluctuations in the film showing small changes in the
       striped domain state. Large variations occur close to the
       edges in the flux closure domains.}}
  \end{figure}

Having calculated the theoretical values for $\omega_{\text{MAG}}$ and $\omega_{\text{DWR}}$, 
we simulate the magnetization dynamics in the film, where we introduce thermal fluctuations due to room temperature. 
Mumax3 simulates finite temperature via a fluctuating thermal field
\cite{brown1963thermal}, which is applied over the film and the system
is allowed to relax. The simulation is run for $20\,\text{ns}$ which
was enough for the dynamics to reach steady state. 
At room temperature, we observe deviations in the magnetization
of the film from its ground state as shown in Fig. \ref{fig:Fig-7}. 
The domain walls between the stripe domains remain intact in
almost the entire cross section of the film. Also, the highest fluctuations
are observed in the flux closure domains near the edges, which form a
periodic pattern. We then vary the magnetic fields applied along the easy axis
and observe their effect on the precession frequency of magnons in the BLIG film.
\subsection{Brillouin light scattering spectroscopy (BLS)}\label{BLSS}
\par The inelastic scattering of photons by elementary excitations
has proved to be a powerful tool for the study of magnetization dynamics \cite{Sebastian15}, \cite{Srinivasan87}, \cite{Demokritov06}.
Photons scattered with higher (lower) energy than that of the incident one constitute anti-Stokes (Stokes) component of BLS.
  \begin{figure}[th]
   \centering{} {
     \includegraphics[width=0.8\columnwidth]{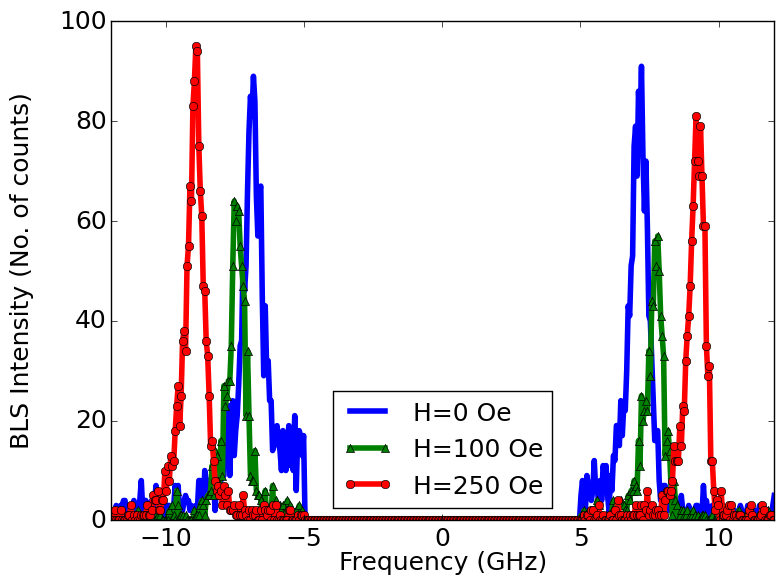}
     \protect\caption{\label{fig:Fig-8}: BLS spectra for
       varying magnetic fields showing the Stokes and anti-Stokes
       components.}}
  \end{figure}
Our BLS spectrometer consists of a tandem Fabry-Perot interferometer and a laser source
of wavelength 532 nm and average power 400 mW. To prevent any local heating by the laser, 
the incident optical power on the film is brought down to 50 mW across a spot of size 30
- 40 $\mu$m using external optics. We observe the BLS scattering peak for a zero applied 
magnetic field at approximately 7 GHz which is of the order of the theoretically calculated
value for the oscillations of thermal magnons in Fig.~\ref{fig:Fig-8}. We then vary the magnetic field and observe 
its effect on the thermal magnon spectra. The obtained spectra in the figure
shows two sets of peaks corresponding to the Stokes and the anti-Stokes components. 
The frequency range of our BLS spectrometer has a lower limit of 1 GHz. As this is 
much above the theoretically calculated $\omega_{\text{DWR}}$, we could not record 
domain wall oscillations. The linewidth of the Brillouin peaks is of the order of 
$400\,\text{MHz}$ which is indicative of the low damping present in these materials.

\subsection{Comparison of simulation and experimental studies} 

\begin{figure}[th]
   \centering{} {
    \includegraphics[width=0.85\columnwidth]{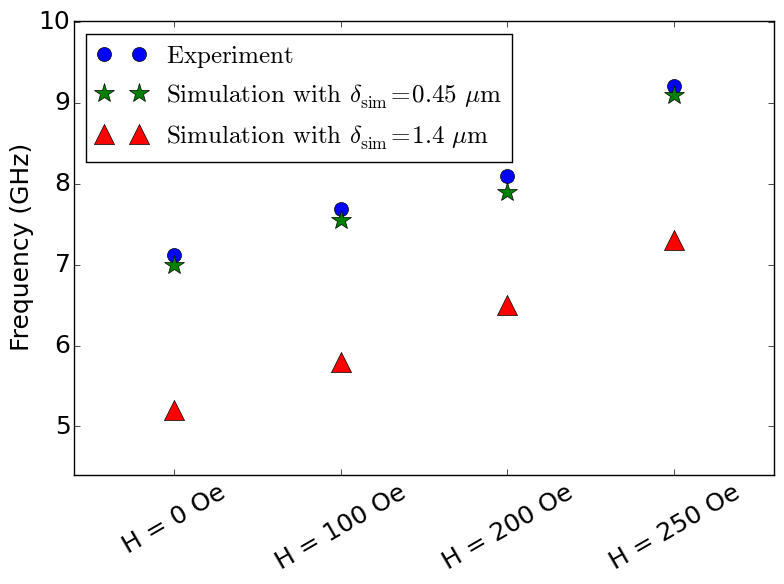}
         \protect\caption{\label{fig:Fig-9} Variation of the thermal magnon
       frequency with applied field obtained from the simulations and
       BLS experiments.}}
  \end{figure}

The oscillation frequencies predicted by the simulations for varying domain widths ($\delta_{\text{sim}}$) of 1.4 $\mu$m and 0.45 $\mu$m and
those observed experimentally in BLS for different applied magnetic fields are compared in Fig.~\ref{fig:Fig-9}. At zero applied field,
when we allow the multi domain simulations to relax with a $\delta_{\text{sim}}$ of 1.4 $\mu$m the oscillation frequencies differ from
the observed experimental values by almost 25\%. The deviation with experiments reduces to less than 2\% when 
we use domains with $\delta_{\text{sim}} \sim 0.45\,\mu\text{m}$. The source of mismatch between simulation and
experiment could be our use of nominal values for material parameters ($M_{\rm s},\,K\,\text{and}\,A$) and film
thickness in our simulations. For the present purpose, it is instructive to note that the quadratic increase in 
magnon frequency, for either domain width, matches the experimental observations.
\section{Summary}
\par In this work, we have demonstrated how to estimate the parameters of a given magnetic film.
For the purpose of obtaining MH curves, we developed an MOFE magnetometer, as an alternative to the 
conventionally used vibrating sample magnetometer.
Polarization microscopy studies allowed us to extract domain widths and observe 
domain patterns for varying magnetic fields. 
\par We know that for any magnetic film, the initial state is multi domain and must be accounted for 
in simulations. We prove and experimentally validate with BLS, 
the ability of MuMax3 for multi domain simulations. These results help improve our
understanding and aid us in building functional devices using any magnetic film. It is possible to obtain a good match between
simulations and experiments with a good knowledge of the material parameters of the film.
\section*{Acknowledgements}
We would like to acknowledge the support of Dr. Ananth Krishnan, IITM, Chennai, for allowing
us to use the optical microscope for imaging of the magnetic domains. We would also like to thank the High Performance Computing
Centre at IIT Madras for the use of their GPU cluster for running our
simulations, and the Dept. of Science and Technology, India for
financial support under Sanction No. SB/S3/EECE/011/2014(IITM).

\section*{References}
\biboptions{numbers,sort&compress}
\bibliographystyle{elsarticle-num}
\bibliography{icmagma}

\end{document}